\documentclass[runningheads]{llncs}
\usepackage{graphicx}

\usepackage[utf8]{inputenc}
\usepackage{csquotes}
\usepackage{appendix}
\usepackage{amsmath}
\usepackage{amssymb}
\usepackage{amsfonts}
\usepackage{hyperref}
\usepackage{tikz-cd}
\usepackage{amsthm}
\usepackage{algorithmic}
\usepackage[ruled,vlined]{algorithm2e}
\usepackage{tabu}
\usepackage{float}
\theoremstyle{definition}
\usepackage{mathtools}

\theoremstyle{plain}

\theoremstyle{definition}

\newcommand{\thistheoremname}{}
\newtheorem*{genericthm*}{\thistheoremname}
\newenvironment{namedthm*}[1]
  {\renewcommand{\thistheoremname}{#1}%
   \begin{genericthm*}}
  {\end{genericthm*}}

\usepackage[backend=bibtex, maxbibnames = 20]{biblatex}
\usepackage[utf8]{inputenc}
\usepackage{graphicx}
\usepackage{amsmath, enumerate, amssymb, bbm, mathtools}
\usepackage{booktabs}
\usepackage{amsthm}
\usepackage{bbm}
\usepackage{multirow}
\usepackage{tikz-cd}
\usepackage[english]{babel}
\usepackage{todonotes}

\begin{document}
\title{Jolt Atlas: Verifiable Inference via Lookup Arguments in Zero Knowledge}
\author{Wyatt Benno, Alberto Centelles, Antoine Douchet, \\ Khalil Gibran}
\authorrunning{W. Benno, A. Centelles, A. Douchet, K. Gibran}
\institute{ICME Labs}
\maketitle

\begin{abstract}
   We present Jolt Atlas, a zero-knowledge machine learning (zkML) framework that extends the Jolt proving system to model inference. Unlike zkVMs (zero-knowledge virtual machines), which emulate CPU instruction execution, Jolt Atlas adapts Jolt's lookup-centric approach and applies it directly to ONNX tensor operations. The ONNX computational model eliminates the need for CPU registers and simplifies memory consistency verification. In addition, ONNX is an open-source, portable format, which makes it easy to share and deploy models across different frameworks, hardware platforms, and runtime environments without requiring framework-specific conversions.
    Our lookup arguments, which use sumcheck protocol, are well-suited for non-linear functions—key building blocks in modern ML. We apply optimisations such as neural teleportation to reduce the size of lookup tables while preserving model accuracy, as well as several tensor-level verification optimisations detailed in this paper. We demonstrate that Jolt Atlas can prove model inference in memory-constrained environments—a prover property commonly referred to as *streaming*. Furthermore, we discuss how Jolt Atlas achieves zero-knowledge through the BlindFold technique, as introduced in Vega. In contrast to existing zkML frameworks, we show practical proving times for classification, embedding, automated reasoning, and small language models.
    Jolt Atlas enables cryptographic verification that can be run on-device, without specialised hardware. The resulting proofs are succinctly verifiable. This makes Jolt Atlas well-suited for privacy-centric and adversarial environments. In a companion work, we outline various use cases of Jolt Atlas, including how it serves as guardrails in agentic commerce and for trustless AI context (often referred to as \textit{AI memory}).
\end{abstract}

\keywords{zkML, Jolt, sumcheck, ONNX, verifiable AI, zero-knowledge, lookup arguments, neural teleportation, agentic commerce}

\newpage
\tableofcontents
\newpage

\section{Introduction}

At the intersection of machine learning, cryptography, and computer science, an active and very lively field is emerging called zero-knowledge machine learning (zkML). It enables verifiable inference with potentially hidden model weights, hidden input data, and even hidden model architecture. zkML protocols build on the techniques from the more general and long-studied field of zero-knowledge succinct arguments of knowledge (zkSNARKs). These allow a prover to convince a verifier that a statement is true, without revealing any information beyond the statement's validity. Such arguments are often succinctly verifiable, meaning that even though the underlying program and prover computation can take significant time and computational resources, the computational work of the verifier is cheap. In the case of zkML, the statements being proven are about machine learning models.

Historically, the task of achieving practical zkML systems has been considered highly impractical. This is due to a combination of factors. Modern neural networks comprise millions to billions of parameters. Even evaluating the model entails substantial arithmetic on high-dimensional tensors. General-purpose SNARK proving still imposes a substantial 100,000X to 1,000,000X overhead over native execution ~\cite{a16zbigideas}. 

The situation is compounded by the fact that neural networks involve more than linear algebra; they rely heavily on non-linear operations such as softmax or ReLU whose faithful realisation in low-degree arithmetic constraint systems can blow up constraint counts and/or degrees, often making them the dominant cost unless specialized techniques such as lookup arguments are used.  Meanwhile, even the ostensibly “friendly” parts of ML, namely matrix multiplication and tensor contractions, dominate FLOPs and arise at large scale; proving them efficiently requires protocols that exploit their algebraic structure, because treating them as generic circuits tends to introduce at minimum large constant-factor overheads.

Finally, many SNARK constructions require materialising large portions of the witness and intermediate polynomials in memory to produce commitments and openings. For ML inference, this can translate into multi-gigabyte peak memory even for modest models, pushing beyond the RAM budget of commodity or on-device provers.

This work addresses these challenges by adapting the lookup-based proving paradigm from Jolt~\cite{jolt} to the computational model of neural networks, moving from CPU instruction execution to tensor computation.

Jolt is a zkVM that achieves state-of-the-art proving performance for RISC-V programs through a novel combination of lookup arguments and the sumcheck protocol. Lookups eliminate the need to arithmetise non-linear functions, while sumcheck verifies large batches of computations with logarithmic verifier work and enables streaming evaluation. Exploiting the structure of computation when applying the sumcheck protocol is central to Jolt's efficiency~\cite{sumcheckisallyouneed}. Their techniques and implementation are thoroughly documented~\cite{joltdocs}.

Below we summarise the core ideas behind Jolt that are adopted in this work.

\subsection{Jolt's inheritance}

\paragraph{Lookup-centric proving.}
A central design choice in Jolt is to express instructions as \emph{lookups} rather than as arithmetic constraints.
Instead of proving complex relations directly as polynomials, the prover reduces to membership statements of the form ``$(q,v)$ appears in a table $\mathcal{T}$,'' where $\mathcal{T}$ captures an operation's behavior.
This is particularly effective for non-linear and piecewise primitives (comparisons, bit-level behavior, range checks) that are expensive in standard circuit encodings.
This lookup-first approach is precisely the mechanism Jolt Atlas adopts for ML activations and other non-linear tensor operators (Section~\ref{sec:lookups}).

\paragraph{Proof as a DAG of sumchecks (staging and batching).}
Jolt organises the proof as a directed acyclic graph (DAG) of sumcheck instances: nodes are sumchecks, edges are evaluation claims that feed later checks.
This dependency structure determines which sumchecks can be batched together and which must be staged sequentially. Jolt Atlas inherits this viewpoint almost verbatim.

\paragraph{Twist and Shout: fast memory checking.}
Jolt's ``Twist and Shout'' optimisation accelerates memory consistency checking using one-hot addressing and structured increments that reduce expensive bookkeeping in RAM/register trace verification.
Although ONNX computation removes general-purpose registers and arbitrary RAM access, we still need a notion of \emph{consistency of reads and writes} over tensor buffers (Section~\ref{sec:architecture}); the lesson from Twist and Shout is that domain-specific access patterns admit much cheaper consistency proofs than prior offline memory checking arguments~\cite{BLUM94}.

\paragraph{Virtual polynomials.}
Jolt avoids committing to large sparse polynomials by defining them \emph{virtually} as algebraic combinations of a small set of low-degree polynomials. A virtual polynomial is a part of the witness that is never committed directly. The term ``virtual polynomial'' was originally introduced in the Binius paper~\cite{binius}.

\paragraph{Prefix-suffix decomposition and streaming (small-space proving).}
Jolt's lookup arguments can be implemented in small space by algebraically decomposing large tables so that their multilinear extension factors into prefix and suffix components evaluated on smaller domains~\cite{NTM25}.
This enables streaming provers that trade time (typically $O(C)$ streaming passes) for space, reducing peak memory from $O(|\mathcal{T}|)$ to roughly $O(|\mathcal{T}|^{1/C})$ while incurring total prover time on the order of $O\!\left(C|\mathcal{T}|\right)$ field operations, where $\mathcal{T}$ is the number of non-zero elements of a table, for a tunable parameter $C$.
This directly motivates Jolt Atlas's focus on ``streaming'' proving for large models on constrained devices (Section~\ref{sec:lookups}).

\paragraph{Zero knowledge via BlindFold}: Jolt obtains zero knowledge by applying the BlindFold technique, with small overhead.

\subsection{Our Contributions}

Jolt Atlas extends the Jolt paradigm to machine learning by replacing the RISC-V instruction set architecture with ONNX computational graphs. This substitution yields several architectural differences from the Jolt zkVM:

\begin{enumerate}
    \item \textbf{DAG-structured computation}: ONNX models are directed acyclic graphs with deterministic data flow, enabling optimisations not available in general-purpose computation. Additionally, ONNX graphs do not use CPU registers or RAM.

    \item \textbf{Tensor operations}: Jolt Atlas operates on tensors (i.e., multi-dimensional vectors of scalars), rather than scalars. Each ONNX operation reads from specific input tensors and writes to a designated output tensor in a pattern known at preprocessing time. Instead of naively decomposing tensor operations into scalar computations and verifying each element independently, we verify tensor relations directly at the multilinear polynomial level.

    \item \textbf{Faster commitments}: Unlike Jolt, our computational model doesn't require us to commit to the full witness at once, but rather to the inner values of tensors. Dory, the polynomial commitment scheme used in Jolt, is only justified for witnesses over $2^{26}$ elements. We therefore replace it with HyperKZG~\cite{hyperkzg}, a KZG-based PCS for multilinear polynomials derived from the Gemini transformation~\cite{gemini}, trading transparent setup for pairing-based succinct openings that are well suited to on-chain verification. 
\end{enumerate}

In our companion work ``Agentic Trust: Solving the Principal-Agent
Problem in Autonomous Commerce Through
Cryptographic Verification", we show how to use zkML with Automated Reasoning models for succinctly verifiable agentic guardrails with over 99\% accuracy.

\subsection{Related Work}

\textbf{EZKL}~\cite{ezkl} is a zkML framework built on the Halo2 proof system from ONNX that relies on both circuit-based arithmetic constraint systems. As already mentioned, this incurs high costs for non-linear operations. While they introduced lookup tables (meaning roughly two advice columns, and two fixed columns of the Halo2 “grid”) for certain non-linear operations such as ReLU, these tables cannot be materialised for large ranges, so they are restricted to small-value operations. Furthermore, the Halo2 proof system does not natively exploit sparsity in lookups or in the matrix-multiplication structure, so the prover pays a largely dense constraint cost.

\textbf{Bionetta}~\cite{bionetta} is a Groth16-based~\cite{groth16} zkML framework focused on client-side proving and zero-knowledge for EVM on-chain verification. As a variant of Groth16, UltraGroth also requires a trusted setup. In Bionetta, weights are hardcoded into the circuit during compilation/setup. That lets them implement matrix-vector and matrix-matrix multiplications with a fixed matrix in 0 constraints (note that this doesn't mean that general MatMul is free). However, like EZKL, it suffers from the fundamental limitation that arithmetic circuits are inefficient for non-linear operations. Furthermore, it doesn't support general ONNX models.

\textbf{DeepProve}~\cite{deepprove} uses a GKR-style sumcheck-based proof system based on the Ceno zkVM~\cite{ceno}. As described in~\cite{thaler2022proofs}, the GKR protocol is considered a general-purpose technique and "general-purpose techniques should sometimes be viewed as heavy hammers that are capable of pounding arbitrary nails, but are not necessarily the most efficient way of hammering any particular nail". In particular, for matrix multiplication, a primitive operation in ML models, "the GKR protocol introduces at least a constant factor overhead for the prover. In practice, this is the difference between a prover that runs many times slower than an (unverifiable) MATMULT algorithm, and a prover that runs a fraction of a percent slower". Furthermore, operations such as Gather require reading the tensor on which they operate, so we need to prove that those reads were correct. The most recent open-source implementation of DeepProve lacks these lookup arguments.

\section{Jolt Atlas Architecture}
\label{sec:architecture}

We now describe how Jolt Atlas organises its zero-knowledge proof of an ONNX model as a DAG of sumcheck instances that feed into a succinct R1CS verifier, using BlindFold to hide the witness.

\subsection{Execution Trace}

An ONNX model execution produces a \emph{trace}: a sequence of computational steps where each step corresponds to one ONNX operation applied to specific tensor elements. Each trace entry contains:
\begin{itemize}
    \item The operation type (Add, Mul, ReLU, etc.)
    \item Source tensor addresses and values
    \item Destination tensor address and value
    \item For non-deterministic operations (e.g., division), an advice value
\end{itemize}

The trace length $T$ determines the constraint system size. Traces are padded to the next power of two for efficient polynomial representation.

\subsection{Proof DAG Structure}

The proof is organised as a DAG where nodes represent sumcheck instances and edges represent polynomial evaluation claims passed between stages. Two sumchecks cannot be batched together if one depends on the other's output. This dependency structure defines the staging.

Following Jolt's architecture~\cite{joltdocs}, proving proceeds through multiple stages. Each stage produces sumcheck proofs with hiding commitments, and the round polynomial coefficients become witness elements in the succinct verifier R1CS.

The terminology in the coming sections follows the excellent survey ``Sum-check is all you need"~\cite{sumcheckisallyouneed}.

\textbf{Stage 1: Outer Sumcheck.} The \texttt{SpartanDag} proves the outer R1CS constraint:
\[
\sum_x \widetilde{eq}(\tau, x) \cdot (\tilde{A}z(x) \cdot \tilde{B}z(x) - \tilde{C}z(x)) = 0,
\]
producing a random point $r = (r_{\text{cycle}}, r_{\text{var}})$ and committed claims $\tilde{A}z(r), \tilde{B}z(r), \tilde{C}z(r)$.

\textbf{Stage 2: Inner Sumcheck and Virtualisation.} 

\emph{Inner sumcheck}. The verifier samples a random linear combination coefficient $\rho$ and the prover shows:
\[
\tilde{A}z(r) + \rho\,\tilde{B}z(r) + \rho^2\,\tilde{C}z(r) \;=\; \sum_y \bigl(\tilde{A}_{\text{small}}(r_{\text{var}}, y) + \rho\,\tilde{B}_{\text{small}}(r_{\text{var}}, y) + \rho^2\,\tilde{C}_{\text{small}}(r_{\text{var}}, y)\bigr) \cdot z(y)
\]
where $z(y)$ is the vector whose entries are the evaluations of the witness polynomial $\widetilde{P}_i(r_{\text{cycle}})$.

\emph{Product virtualisation sumchecks}. Witness variables defined as element-wise products of two committed polynomials:
\[
\widetilde{V}(r_{\text{cycle}}) \;=\; \sum_t \widetilde{eq}(r_{\text{cycle}}, t) \cdot L(t) \cdot R(t).
\]
The four products virtualised are:
\begin{itemize}
    \item \text{Combined instruction lookup input}: $\text{LeftInput}(t) \times \text{RightInput}(t)$.
    \item \text{Selective write to destination tensor}: $\text{td\_addr}(t) \times \text{write\_flag}(t)$ 
    \item \text{Conditional selection condition}: $\text{ts1\_val}(t) \times \text{select\_flag}(t)$ 
    \item \text{Conditional selection result}: $\text{td\_write\_val}(t) \times \text{select\_flag}(t)$ 
\end{itemize}

\textbf{Stage 3: Lookups and Instruction Verification.} Batches:
\begin{itemize}
    \item \texttt{PCSumcheck}: Ensures each execution step (PC) transitions to the next instruction (NextPC):
    \[
      \widetilde{\mathrm{NextPC}}(r_{\mathrm{cycle}})
      \;=\;
      \sum_{t} \widetilde{\mathrm{PC}}(t)\cdot \widetilde{eq}_{+1}(r_{\mathrm{cycle}}, t).
    \]
    \item \texttt{ReadRafSumcheck}: A sparse-dense sumcheck verifying lookup-table reads. Each 64-bit lookup index is decomposed into chunks, and a prefix-suffix protocol checks:
    \[
      \widetilde{\mathrm{LookupOutput}}(r_{\mathrm{cycle}})
      \;=\;
      \sum_{k,\,j} \widetilde{eq}(r_{\mathrm{addr}}, k)\cdot\widetilde{eq}(r_{\mathrm{cycle}}, j)\cdot\mathrm{val}(k)\cdot\mathrm{ra}(k),
    \]
      where $\mathrm{ra}(k)$ is the random-address indicator polynomial and $\mathrm{val}(k)$ the table value polynomial. The prefix-suffix decomposition reduces the prover's memory footprint from $O(|\mathcal{T}|)$ to $O(|\mathcal{T}|^{1/C})$.
    \item \texttt{BooleanitySumcheck}: Verifies that each $\mathrm{ra}_i$ chunk polynomial (the random-address accumulations) satisfies the booleanity constraint
    \[
      \mathrm{ra}_i(k)^2 - \mathrm{ra}_i(k) = 0
    \]
    \item \texttt{HammingWeightSumcheck}: Verifies that each $\mathrm{ra}_i$ chunk polynomial is a one-hot vector. The claimed sum is
    \[
      \sum_{k}\sum_{i=0}^{D-1} \gamma^{i}\,\mathrm{ra}_i(k) \;=\; \sum_{i=0}^{D-1} \gamma^{i},
    \]
    where $\gamma$ is a fresh Fiat--Shamir challenge. Since the booleanity sumcheck already ensures $\mathrm{ra}_i(k)\in\{0,1\}$, this equality forces $\sum_k \mathrm{ra}_i(k) = 1$ for every chunk~$i$.
    \item \texttt{InstructionInputSumcheck}: Verifies instruction inputs match claimed values
\end{itemize}

\textbf{Stage 4: Random-Address Virtualisation and Memory Checking.} This stage batches two sumchecks:
\begin{itemize}
    \item \texttt{RASumCheck}: Virtualises the $D=8$ per-chunk $\mathrm{ra}_i$ polynomials into $d$ groups, verifying that the full random-address product $\prod_i \mathrm{ra}_i$ equals the claimed evaluation.
    \item \texttt{MemoryDag}: Verifies read-write consistency, ensuring tensor values read during execution match previously written values.
\end{itemize}

\textbf{Stage 5: Memory Value Evaluation.} Verifies the evaluation of the memory-value polynomial $\widetilde{\mathrm{Val}}$ at the random point $(r_{\mathrm{addr}}, r_{\mathrm{cycle}})$ produced by Stage~4's read-write checking.
    The claimed sum $\widetilde{\mathrm{Val}}(r_{\mathrm{addr}}, r_{\mathrm{cycle}})$ is reduced via sumcheck over the cycle dimension:
    \[
      \widetilde{\mathrm{Val}}(r_{\mathrm{addr}}, r_{\mathrm{cycle}})
      \;=\;
      \sum_{j} \mathrm{Inc}(j)\cdot\mathrm{Wa}(j)\cdot\mathrm{Lt}(j, r_{\mathrm{cycle}}),
    \]
here $\mathrm{Inc}(j)$ is the committed increment polynomial, $\mathrm{Wa}(j) = \widetilde{eq}(r_{\mathrm{addr}}, \mathrm{td}(j))$ is the write-allocation polynomial indicating whether cycle~$j$ writes to the queried address, and $\mathrm{Lt}(j, r_{\mathrm{cycle}})$ is the ``less-than'' multilinear extension that selects only writes occurring before the queried cycle.
    The output claims open $\mathrm{Inc}$ and $\mathrm{Wa}$ at the sumcheck's final point.
    
\textbf{Stage 6: Bytecode Verification.} Verifies that executed operations match the committed ONNX graph structure.

\textbf{Stage 7: BlindFold.} Folds the existing instance with a Nova randomised instance.

\section{Zero Knowledge via BlindFold}
\label{sec:zk}

Jolt Atlas' prover executes multiple sumcheck stages (e.g., bytecode, instruction lookups, etc.), each producing a transcript of polynomial coefficients and verifier challenges. These transcripts are not zero-knowledge, since they reveal the witness. BlindFold~\cite{vega} retrofits zero-knowledge onto these proofs by encoding the sumcheck verifier as an R1CS circuit and applying Nova-style folding with a random satisfying pair. Like Jolt, we achieve zero-knowledge via BlindFold by:

\begin{enumerate}
    \item \textbf{Hiding sumcheck messages}: All prover messages (round polynomial coefficients, claimed evaluations) are sent as Pedersen commitments rather than plaintext values.
    \item \textbf{Nova-style folding}: The existing instance is folded with a random satisfying instance, producing a folded witness $w_{\text{folded}} = w + r \cdot w_{\text{rand}}$ that reveals nothing about the original witness.
    \item \textbf{Succinct verifier R1CS}: We construct a small R1CS circuit encoding only the verifier's algebraic checks. 
\end{enumerate}

\subsection{The BlindFold Protocol}

In this section we provide a description of BlindFold as described in~\cite{vega}.

\subsubsection{Relaxed R1CS}

Standard R1CS $(A\mathbf{Z}) \circ (B\mathbf{Z}) = C\mathbf{Z}$ does not fold cleanly due to cross-terms. Following Nova~\cite{KST22}, we use relaxed R1CS:
\begin{equation}
(A\mathbf{Z}) \circ (B\mathbf{Z})
= u \cdot (C\mathbf{Z}) + \mathbf{E},
\end{equation}
where $u \in \mathbb{F}$ is a scalar and $\mathbf{E} \in \mathbb{F}^m$ is an error vector. A standard R1CS instance is a special case with $u=1$ and $\mathbf{E}=\mathbf{0}$.

\paragraph{Instance (public).}
$\mathcal{U} = (\bar{E}, u, \bar{W}, \mathbf{x}, \{\bar{R}_i\}, \{\bar{V}_i\})$
\begin{itemize}
  \item $\bar{E} = \mathrm{Com}(\mathbf{E}, r_E)$: Pedersen commitment to the error vector
  \item $\bar{W} = \mathrm{Com}(\mathbf{W}, r_W)$: Pedersen commitment to the witness
  \item $\bar{R}_i = \mathrm{Com}(\mathbf{c}^{(i)}, \rho_i)$: per-round commitments to polynomial coefficients
  \item $\bar{V}_i$: evaluation commitments for PCS binding constraints
\end{itemize}

\paragraph{Witness (private).}
$\mathcal{W} = (\mathbf{E}, r_E, \mathbf{W}, r_W, \{\mathbf{c}^{(i)}, \rho_i\})$.

\subsubsection{Folding (Nova-style)}

Given two satisfying relaxed R1CS pairs $(\mathcal{U}_1, \mathcal{W}_1)$ and $(\mathcal{U}_2, \mathcal{W}_2)$:

\paragraph{Cross-term computation.}
Compute $\mathbf{T} \in \mathbb{F}^m$:
\begin{equation}
T_i
= (A\mathbf{Z}_1)_i \cdot (B\mathbf{Z}_2)_i
+ (A\mathbf{Z}_2)_i \cdot (B\mathbf{Z}_1)_i
- u_1 (C\mathbf{Z}_2)_i
- u_2 (C\mathbf{Z}_1)_i.
\end{equation}

This arises from expanding $(A\mathbf{Z}') \circ (B\mathbf{Z}')$ where $\mathbf{Z}' = \mathbf{Z}_1 + r\mathbf{Z}_2$:
\begin{align}
(A\mathbf{Z}') \circ (B\mathbf{Z}')
&=
\underbrace{(A\mathbf{Z}_1)\circ(B\mathbf{Z}_1)}_{\text{from } (\mathcal{U}_1,\mathcal{W}_1)}
+ r \cdot \underbrace{\mathbf{T}}_{\text{cross-term}}
+ r^2 \cdot \underbrace{(A\mathbf{Z}_2)\circ(B\mathbf{Z}_2)}_{\text{from } (\mathcal{U}_2,\mathcal{W}_2)}.
\end{align}

\paragraph{Folded instance (verifier-computable).}
\begin{align}
\bar{E}' &= \bar{E}_1 + r \cdot \bar{T} + r^2 \cdot \bar{E}_2, &
u' &= u_1 + r \cdot u_2, \\
\bar{W}' &= \bar{W}_1 + r \cdot \bar{W}_2, &
\mathbf{x}' &= \mathbf{x}_1 + r \cdot \mathbf{x}_2, \\
\bar{R}'_i &= \bar{R}_{1,i} + r \cdot \bar{R}_{2,i}, &
\bar{V}'_i &= \bar{V}_{1,i} + r \cdot \bar{V}_{2,i}.
\end{align}

\paragraph{Folded witness (prover-only).}
\begin{align}
\mathbf{E}' &= \mathbf{E}_1 + r \cdot \mathbf{T} + r^2 \cdot \mathbf{E}_2, &
r'_E &= r_{E_1} + r \cdot r_T + r^2 \cdot r_{E_2}, \\
\mathbf{W}' &= \mathbf{W}_1 + r \cdot \mathbf{W}_2, &
r'_W &= r_{W_1} + r \cdot r_{W_2}.
\end{align}

Correctness follows from Pedersen's additive homomorphism:
\begin{equation}
\mathrm{Com}(\mathbf{W}', r'_W)
= \bar{W}_1 + r \cdot \bar{W}_2
= \bar{W}'.
\end{equation}

\subsubsection{Protocol (Prover)}

\paragraph{Input.}
Real instance-witness pair $(\mathcal{U}_1, \mathcal{W}_1)$ with $u_1=1$ and $\mathbf{E}_1=\mathbf{0}$ (non-relaxed, from actual sumcheck execution).

\begin{enumerate}
  \item Sample random satisfying pair $(\mathcal{U}_2, \mathcal{W}_2, \mathbf{Z}_2)$:
  \begin{itemize}
    \item Sample $\mathbf{W}_2 \xleftarrow{\$} \mathbb{F}_p^n$ with the same structural layout (coefficients, intermediates, next-claims per round).
    \item Sample $u_2 \xleftarrow{\$} \mathbb{F}_p\setminus\{0\}$ and $\mathbf{x}_2 \xleftarrow{\$} \mathbb{F}_p$.
    \item Compute $\mathbf{E}_2 = (A\mathbf{Z}_2)\circ(B\mathbf{Z}_2) - u_2 \cdot (C\mathbf{Z}_2)$ to force satisfaction.
    \item Commit: $\bar{E}_2 = \mathrm{Com}(\mathbf{E}_2, r_{E_2})$, $\bar{W}_2 = \mathrm{Com}(\mathbf{W}_2, r_{W_2})$.
    \item Extract round coefficients from $\mathbf{W}_2$ and commit: $\bar{R}_{2,i} = \mathrm{Com}(\mathbf{c}_2^{(i)}, \rho_{2,i})$.
  \end{itemize}
  \item Compute cross-term $\mathbf{T}$ and commit: $\bar{T} = \mathrm{Com}(\mathbf{T}, r_T)$.
  \item Fiat--Shamir challenge: append $(\mathcal{U}_1, \mathcal{U}_2, \bar{T})$ to the transcript; derive $r \leftarrow \mathcal{H}(\text{transcript})$.
  \item Fold: compute $(\mathcal{U}', \mathcal{W}')$ as explained above.
  \item Output proof: $\pi = (\mathcal{U}_1,\ \mathcal{U}_2,\ \bar{T},\ \mathcal{W}')$.
\end{enumerate}

\subsubsection{Protocol (Verifier)}

\paragraph{Input.}
Proof $\pi = (\mathcal{U}_1, \mathcal{U}_2, \bar{T}, \mathcal{W}')$ and the Fiat--Shamir transcript.

\begin{enumerate}
  \item Check non-relaxed: verify $u_1 = 1$ and $\bar{E}_1 = \mathcal{O}$ (identity element).
  \item Replay Fiat--Shamir: recompute $r$ from $(\mathcal{U}_1, \mathcal{U}_2, \bar{T})$.
  \item Recompute folded instance $\mathcal{U}'$ from public data.
  \item Commitment openings: verify
  \begin{itemize}
    \item $\bar{W}' \stackrel{?}{=} \mathrm{Com}(\mathbf{W}', r'_W)$,
    \item $\bar{E}' \stackrel{?}{=} \mathrm{Com}(\mathbf{E}', r'_E)$,
    \item $\bar{R}'_i \stackrel{?}{=} \mathrm{Com}(\mathbf{c}'^{(i)}, \rho'_i)$ for each round.
  \end{itemize}
  \item Coefficient consistency: verify the coefficients embedded in $\mathbf{W}'$ match $\{\mathbf{c}'^{(i)}\}$ (prevents using different coefficient values in the R1CS witness vs.\ the committed round polynomials).
  \item Evaluation commitment check (PCS binding): for each extra constraint $i$, verify
  \begin{equation}
    \bar{V}'_i = y'_i \cdot G + b'_i \cdot H,
  \end{equation}
  where $y'_i, b'_i$ are extracted from $\mathbf{W}'$.
  \item R1CS satisfaction: verify
  \begin{equation}
    (A\mathbf{Z}') \circ (B\mathbf{Z}')
    = u' \cdot (C\mathbf{Z}')
    + \mathbf{E}'.
  \end{equation}
\end{enumerate}

\subsection{R1CS Encoding of the Sumcheck Verifier}
To apply BlindFold, we need to encode all of the verifier's algebraic checks — across all stages and rounds — into a single R1CS instance $(A\mathbf{Z}) \circ (B\mathbf{Z}) =  C\mathbf{Z}$, so that the BlindFold folding protocol can prove the sumcheck transcripts are valid without revealing them.

Let the Jolt Atlas proof consist of $S$ stages, where stage $s$ has $n_s$ sumcheck rounds with degree-$d_s$ univariate polynomials, and let $N = \sum_s n_s$ denote the total number of rounds. We define the witness vector $\mathbf{Z}$ as:
\begin{equation*}
\mathbf{Z} =
\Bigl[
\underbrace{u}_{\text{scalar}},\ 
\underbrace{r_1,\ldots,r_N,\ \sigma_0^{(1)},\ldots,\sigma_0^{(k)},\ \alpha_1,\ldots,\gamma_1,\ldots}_{\text{public inputs } \mathbf{x}},\ 
\underbrace{c_0^{(1)},c_1^{(1)},\ldots,t_1^{(1)},\ldots,\sigma_1^{(1)},\ldots}_{\text{private witness } \mathbf{W}}
\Bigr],
\end{equation*}

The public input region contains values known to both
  prover and verifier: the $N$ Fiat-Shamir challenges
  $r_j$, the $K$ initial claimed sums $\sigma_0^{(k)}$
  (one per independent chain), batching coefficients
  $\alpha_j$, and constraint challenge values
  $\gamma_i$. The private witness region contains, for
  each round $j$, the polynomial coefficients
  $(c_0^{(j)}, \ldots, c_{d_s}^{(j)})$, the $d_s - 1$
  Horner intermediates $(t_0^{(j)}, \ldots,
  t_{d_s-2}^{(j)})$, and the next claimed sum
  $\sigma_{j+1}$. Additional witness slots hold
  polynomial evaluation values $y_{\omega_k}$ and
  auxiliary variables for output/input binding
  constraints (described below).

For each round $j$ of stage $s$, the sumcheck prover sends coefficients $(c_0, c_1, \ldots, c_{d_s})$ and receives a challenge $r_j \xleftarrow{\$} \mathbb{F}$. The verifier checks two relations:

\begin{enumerate}
    \item \textit{Sumcheck identity}:
    \begin{equation*}
        2c_0 + c_1 + c_2 + \cdots + c_{d_s} = \sigma_j,
    \end{equation*}
    where $\sigma_j$ is the claimed sum (output of the previous round, or the initial claim $\sigma_0$ for the first round).
    \item Horner evaluation (next claimed sum).
Let $g_j(X) = \sum_{\ell=0}^{d_s} c_\ell X^\ell$ be the prover's degree-$d_s$ univariate polynomial in round $j$.
The verifier checks that the next claimed sum satisfies
\begin{equation}
\sigma_{j+1} \;=\; g_j(r_j)
\;=\; c_0 + r_j\!\left(c_1 + r_j\!\left(c_2 + \cdots + r_j c_{d_s}\right)\right).
\end{equation}

To encode this efficiently in R1CS, we introduce auxiliary variables
$t_0,\ldots,t_{d_s-2}$ implementing Horner's rule:
\begin{align}
t_{d_s-2} &:= c_{d_s-1} + r_j \cdot c_{d_s}, \label{eq:horner_last}\\
t_{i-1} &:= c_i + r_j \cdot t_i \qquad \text{for } i = d_s-2, d_s-3, \ldots, 1, \label{eq:horner_step}\\
\sigma_{j+1} &:= c_0 + r_j \cdot t_0. \label{eq:horner_out}
\end{align}

Each multiplication in~\eqref{eq:horner_last}--\eqref{eq:horner_out} is enforced by one multiplicative R1CS constraint
of the form $\langle A, \mathbf{Z}\rangle \cdot \langle B, \mathbf{Z}\rangle = \langle C, \mathbf{Z}\rangle$:
\begin{align}
\text{(innermost)}\quad &\langle A,\mathbf{Z}\rangle = c_{d_s},\quad
\langle B,\mathbf{Z}\rangle = r_j,\quad
\langle C,\mathbf{Z}\rangle = t_{d_s-2} - c_{d_s-1}, \label{eq:r1cs_innermost}\\
\text{(middle)}\quad &\langle A,\mathbf{Z}\rangle = t_i,\quad
\langle B,\mathbf{Z}\rangle = r_j,\quad
\langle C,\mathbf{Z}\rangle = t_{i-1} - c_i,
\qquad i = d_s-2,\ldots,1, \label{eq:r1cs_middle}\\
\text{(final)}\quad &\langle A,\mathbf{Z}\rangle = t_0,\quad
\langle B,\mathbf{Z}\rangle = r_j,\quad
\langle C,\mathbf{Z}\rangle = \sigma_{j+1} - c_0. \label{eq:r1cs_final}
\end{align}

Equivalently, \eqref{eq:r1cs_innermost}--\eqref{eq:r1cs_final} enforce exactly the Horner recurrences
\eqref{eq:horner_last}--\eqref{eq:horner_out}.
\end{enumerate}

\paragraph{Circuit size.}
Each sumcheck round contributes one sumcheck-identity constraint and $d_s$ Horner-evaluation constraints, for a total of $d_s + 1$ constraints per round. Output and input-binding constraints add a number of gates proportional to the number of terms in the sum-of-products expression, which is bounded by a constant per stage. The total R1CS size is therefore
\begin{equation}
O\!\left(\sum_s n_s \cdot d_s\right)
\;=\;
O\!\left(N \cdot d_{\max}\right),
\end{equation}
where $N$ is the total number of sumcheck rounds across all stages and $d_{\max}$ is the maximum polynomial degree. Since the number of sumcheck rounds is logarithmic in the original computation size (each round halves the domain), the BlindFold circuit is logarithmic in the size of the computation being proven.

\subsection{Integration with Jolt Atlas}

During stages prior to BlindFold, each sumcheck instance records the following data into an \emph{opening accumulator} (a running collection of proof artifacts):
\begin{itemize}
  \item Polynomial coefficients $\{c_0^{(j)}, c_1^{(j)}, \ldots, c_d^{(j)}\}$ for each round~$j$, representing the prover's univariate message $g_j(X)$.
  \item Round commitments $\bar{R}_j = Com(\mathbf{c}^{(j)}, \rho_j)$: Pedersen commitments to each round's coefficients, sent to the verifier during the sumcheck.
  \item Blinding factors $\rho_j$: the randomness used in each round commitment.
  \item Verifier challenges $r_j$: the Fiat--Shamir challenges derived after each round.
  \item Initial claimed sum $\sigma_0^{(s)}$ for each stage~$s$.
  \item Input/output claim constraints: the algebraic relations that link each stage's initial claim (resp.\ final output) to polynomial evaluations from other stages (as described in the input/output binding sections).
  \item Polynomial evaluation values $y_{\omega_k}$: retrieved from the accumulator by opening identifier $\omega_k$.
\end{itemize}


From the extracted data, the prover constructs the BlindFold R1CS as follows:
\begin{enumerate}
  \item \textbf{Stage configuration.}
  Each sumcheck round becomes one R1CS ``stage'' with its polynomial degree, chain linkage information, and (for the first and last rounds of each Jolt Atlas stage) input/output constraints.

  \item \textbf{Constraint assembly.}
  The builder processes all stage configurations in sequence. For each stage it allocates private
  witness variables (coefficients, Horner intermediates, next-claim), emits the sum-check identity constraint and the Horner evaluation gates, and chains the round's output variable to the next round's input variable. 
  Input and output constraints are attached to the first and last rounds of each Jolt Atlas stage respectively, with polynomial evaluations wired through the global opening map~$\Phi$.

  \item \textbf{PCS binding constraint.}
  After all sumcheck stages, one additional constraint ties the sumcheck reductions to the polynomial commitment scheme. During the joint opening phase, all polynomial evaluations ${y_{\omega_i}}$ accumulated across stages are batched into a single claim:
  \begin{equation*}
    y_{\mathrm{joint}} \;=\; \sum_i \beta_i \cdot y_{\omega_i}
  \end{equation*}
  where $\beta_i$ are batching coefficients derived from the Fiat-Shamir transcript. The PCS proof produces an evaluation commitment $\bar{V} = y_{\text{joint}} \cdot G + b \cdot H$, where
  $b$ is a blinding factor and $G, H$ are public generators. 
  
  The extra R1CS constraint asserts the linear relation above, using the same witness variables $y_{\omega_i}$ that appear in the stage output/input constraints, while $\bar{V}$ is included in the relaxed R1CS instance so that the BlindFold verifier can check that the committed value matches the witness. This is the constraint that binds the sumcheck reductions to the polynomial commitment proof:
  without it, a prover could satisfy all sumcheck constraints with evaluation values that are inconsistent with the committed polynomials.

  \item \textbf{Witness assignment.}
  The prover populates the full witness vector~$\mathbf{Z}$ by placing the initial claims and all Fiat--Shamir challenges into the public-input slots, and filling the private-witness slots with round coefficients, Horner intermediates, next-claims (computed via Horner evaluation), opening values, and PCS blinding factors.

  \item \textbf{Instance construction.}
  A non-relaxed relaxed-R1CS instance $(\mathcal{U}_1,\mathcal{W}_1)$ (i.e., with $u = 1$ and $E = 0$) is formed with $u=1$, $\mathbf{E}=\mathbf{0}$, the witness commitment $\bar{W}=Com(\mathbf{W},r_W)$, all round commitments $\{\bar{R}_j\}$, the evaluation commitment $\bar{V}$ from HyperKZG, and the public inputs.
  Pedersen generators are derived deterministically so that both prover and verifier use the same basis.

  \item \textbf{BlindFold execution.}
  The prover runs the BlindFold protocol (Section~5) on $(\mathcal{U}_1,\mathcal{W}_1)$, producing a proof $\pi$ that is appended to the Jolt Atlas proof.
\end{enumerate}

\paragraph{Verification.}
The verifier reconstructs the same R1CS from the stage configurations (which are deterministic from the circuit structure and the Fiat--Shamir transcript). It then:
\begin{enumerate}
  \item Recovers the public inputs (the sumcheck challenges $\{r_j\}$ and initial claims $\{\sigma_0^{(s)}\}$) by replaying the Fiat--Shamir transcript from the main Jolt Atlas proof.
  \item Reconstructs the real instance $\mathcal{U}_1$ by collecting round commitments $\{\bar{R}_j\}$ from the sumcheck proofs (which are part of the Jolt Atlas proof) and the evaluation commitment $\bar{V}$ from the HyperKZG proof.
  \item Runs the BlindFold verifier on $\pi$, which checks commitment openings, coefficient consistency, and relaxed-R1CS satisfaction on the folded instance.
\end{enumerate}

The verifier never sees the polynomial coefficients, Horner intermediates, or evaluation values, only their commitments. BlindFold guarantees that these hidden values satisfy the sumcheck verification equations, completing the zero-knowledge property.

\section{Lookup Arguments in Small Space}
\label{sec:lookups}

Non-linear operations like Softmax cannot be expressed as polynomial relations. Jolt Atlas uses lookup arguments to verify these operations: the prover demonstrates that each input-output pair appears in a precomputed lookup table.

In essence, a lookup table $\mathcal{T}: \{0,1\}^w \to \mathbb{F}$ maps $w$-bit inputs to field elements. Its multilinear extension $\tilde{\mathcal{T}}$ can be evaluated at any point $r \in \mathbb{F}^w$.

To verify a lookup $(q, v)$ where $v = \mathcal{T}[q]$, the prover commits to the query vector and demonstrates that:
\[
\tilde{\mathcal{T}}(r) = v \quad \text{at random } r
\]
using the sumcheck protocol.

\subsection{Prefix-Suffix Decomposition of Large Lookup Tables}

A naive lookup table for operations on two 64-bit inputs would require $2^{128}$ entries, which is infeasible. Jolt~\cite{jolt} exploits the observation that many operations can be evaluated \emph{chunk-by-chunk}: the result on the full input can be expressed as a function of results on small input segments.

Concretely, the input bits are partitioned into $C$ segments
\(
x = \bigl(x^{(1)}, \ldots, x^{(C)}\bigr),
\)
each of width $b$ (e.g., $C=8$ segments of $b=8$ bits). For a table $\mathcal{T}$ admitting this decomposition, the multilinear extension factors as
\begin{equation}
\label{eq:prefix-suffix-lookup}
\widetilde{\mathcal{T}}(x)
\;=\;
\sum_{i} \prod_{j=1}^{C} s_{i,j}\!\bigl(x^{(j)}\bigr),
\end{equation}
where each subtable function $s_{i,j}$ operates on only $b$ bits, reducing the total table size from $2^{Cb}$ to $O\!\left(C \cdot 2^b\right)$. 

For a detailed treatment of which operations admit this decomposition and the resulting prover complexity, see Appendix~A of~\cite{NTM25}.

\textbf{Example: ReLU.} For a $w$-bit signed integer with most significant bit $b_0$, the ReLU function $\text{ReLU}(x) = \max(0, x)$ is entirely determined by the sign bit:
\[
\text{ReLU}(x) = (1 - b_0) \cdot x.
\]
Split the input into high bits $x_{\text{hi}}$ (containing $b_0$) and low bits $x_{\text{lo}}$, each of width $w/2$. Since $x = \mathrm{val}(x_{\text{hi}}) \cdot 2^{w/2} + \mathrm{val}(x_{\text{lo}})$, we can write:

\begin{align*}
\text{ReLU}(x_{\text{hi}}, x_{\text{lo}})
&= (1 - b_0)\bigl(\mathrm{val}(x_{\text{hi}}) \cdot 2^{w/2} + \mathrm{val}(x_{\text{lo}})\bigr) \\
&= \underbrace{(1 - b_0) \cdot \mathrm{val}(x_{\text{hi}}) \cdot 2^{w/2}}_{p_{\text{Relu}}(x_{\text{hi}})} \cdot \underbrace{1}_{s_{\text{One}}(x_{\text{lo}})}
\;+\;
\underbrace{(1 - b_0)}_{p_{\text{NotMSB}}(x_{\text{hi}})} \cdot \underbrace{\mathrm{val}(x_{\text{lo}})}_{s_{\text{Relu}}(x_{\text{lo}})}
\end{align*}
The first term captures the high-order contribution (which vanishes for negative inputs), and the second passes through the low-order value gated by the sign indicator. Since this identity holds on all Boolean inputs, it extends to the multilinear extensions, giving the decomposition $\widetilde{\text{ReLU}}(r) = p_{\text{Relu}}(r_{\text{hi}}) \cdot s_{\text{One}}(r_{\text{lo}}) + p_{\text{NotMSB}}(r_{\text{hi}}) \cdot s_{\text{Relu}}(r_{\text{lo}})$.

\subsection{Neural Teleportation for Lookup Table Compression}

Activation functions like \texttt{erf} (used in GELU) and \texttt{tanh} require lookup tables spanning the full input range, which increases the computational costs of both prover and verifier. To mitigate this, Jolt Atlas draws on the idea of \emph{neural teleportation} introduced by TeleSparse~\cite{telesparse}, adapting it into a simplified form suited to the proving setting.

TeleSparse defines a two-sided, per-neuron teleportation: for each neuron $i$ with activation $\sigma$, it applies a scalar $\lambda_i > 0$, dividing the pre-activation input by $\lambda_i$ and multiplying the output by a compensating factor, then rescaling the downstream weights accordingly. Because both the activation input and the subsequent weights are adjusted, the transform is lossless, i.e., the network computes the same function.

In the proving setting of Jolt Atlas' implementation, per-neuron weight rescaling would require re-committing to modified weight tensors and complicates the circuit. Jolt Atlas therefore uses a \emph{global, one-sided} approximation: every pre-activation input to a selected activation is divided by a single factor $\tau > 1$, with \emph{no} compensating output multiplication and \emph{no} weight rescaling. Concretely, the original computation $y = \sigma(x)$ is replaced by $y' = \sigma(x / \tau)$, which is a lossy approximation (i.e., $y' \neq y$ in general) but it is not significant in practice.

The key observation is that saturating activations like $\text{erf}$ and $\tanh$ spend most of their operating range in the flat regions where $\sigma(x) \approx \pm 1$. For any input already in the saturation region, $\sigma(x/\tau) = \sigma(x)$, so the approximation is exact. Error is concentrated in the narrow linear region near the origin, where it is bounded and small relative to the output scale. In typical trained networks, the vast majority of pre-activation values fall in the saturation region, so the global one-sided transform introduces only minor distortion.

With $\tau = 4$, the effective input range shrinks by $4\times$, from $[-R, R]$ to $[-R/4, R/4]$, and the lookup table is bounded by $2^{16}$ entries (sufficient for 16-bit fixed-point activations). Empirical evaluation shows that $\tau = 4$ introduces output differences of less than 55 units on a 128-scale fixed-point representation, acceptable for most inference tasks. The factor $\tau$ can be tuned per-model to balance table size against accuracy.

\section{ONNX: The Bridge to Machine Learning}
\label{sec:onnx-details}

ONNX (Open Neural Network Exchange) serves as the standardised interface between machine learning frameworks and Jolt Atlas's verification system. This section describes ONNX's role and how Jolt Atlas handles its computational model.

\subsection{Why ONNX}

ONNX provides a framework-agnostic representation of neural network computations. Models trained in PyTorch, TensorFlow, or JAX can be exported to ONNX format and verified by Jolt Atlas without modification to the training workflow.

The ONNX format represents computations as directed acyclic graphs where:
\begin{itemize}
    \item \textbf{Nodes} represent operations (Add, MatMul, ReLU, etc.)
    \item \textbf{Edges} represent tensors flowing between operations
    \item \textbf{Attributes} parameterise operations (axis specifications, einsum patterns)
    \item \textbf{Initialisers} store model weights and biases
\end{itemize}

\subsection{Operator Support}

ONNX defines over 180 operators across 19 domains. Jolt Atlas currently supports a subset focused on inference workloads:

\begin{itemize}
    \item \textbf{Elementwise Arithmetic:} \texttt{Add}, \texttt{Sub}, \texttt{Mul}, \texttt{Div}
    \item \textbf{Activations (via lookup tables):} \texttt{Relu}, \texttt{Sigmoid}, \texttt{Softmax}, \texttt{Tanh}, \texttt{Erf}
    \item \textbf{Comparisons:} \texttt{Gte}, \texttt{Eq}
    \item \textbf{Math Functions:} \texttt{Rsqrt} (reciprocal square root)
    \item \textbf{Tensor Operations:}
\begin{itemize}
    \item Shape manipulation: \texttt{Reshape}, \texttt{Broadcast} (automatic, NumPy semantics)
    \item Data movement: \texttt{Gather}
    \item Control flow: \texttt{Select} (conditional element selection)
\end{itemize}
    \item \textbf{Reduction Operations:}
\begin{itemize}
    \item \texttt{ReduceSum} along arbitrary axes
    \item \texttt{ReduceMean} (decomposed internally into \texttt{ReduceSum} + \texttt{Div})
\end{itemize}
    \item \textbf{Tensor Contraction:} \texttt{Einsum} (supports patterns like \texttt{``mk,kn->mn''} (matrix multiply), \texttt{``bmk,bkn->bmn''} (batched multiply), and attention patterns)
\end{itemize}

\subsection{From ONNX Graph to Proof}

The verification pipeline proceeds as follows:

\begin{enumerate}
    \item \textbf{Graph parsing}: The ONNX protobuf is parsed into an internal representation, resolving tensor shapes and operator attributes.

    \item \textbf{Preprocessing}: Static analysis determines memory layout, identifies which operations require lookups versus polynomial verification, and generates the bytecode specification.

    \item \textbf{Trace generation}: The model executes on the input, recording all intermediate tensor values. 

    \item \textbf{Proof generation}: The trace feeds into the Jolt Atlas prover, which generates a zero-knowledge proof.
\end{enumerate}

Our initial focus is to support the operators required for common inference workloads (classification, embedding generation, and transformer attention) rather than the full ONNX specification.

\subsection{CPU trace vs Tensor trace}

Currently, to verify tensor operations such as addition, multiplication, or ReLU, we begin with the ONNX trace, decompose it into a CPU trace, and then feed this representation into the Jolt proof system. While multilinear polynomials provide a natural and expressive representation of tensors, this decomposition step discards much of the structure and algebraic expressiveness inherent in the polynomial formulation.

To preserve this structure, we instead verify tensor relations directly at the polynomial level. Because multilinear polynomials naturally encode tensor computations, we can apply the Schwartz–Zippel lemma to probabilistically check the correctness of a tensor operation without examining each individual element.

The core design philosophy closely follows the principles underlying Jolt—arguably even more directly than Jolt itself. In particular, we reduce commitment costs through virtual polynomials and minimize constraint overhead by relying on lookup arguments. Nearly all verification checks are performed using virtual polynomials, and the approach neither uses nor requires R1CS.

\section{Benchmarks}
\label{sec:benchmarks}

\paragraph{System.}
All benchmarks were run on a MacBook Pro (Apple M3) with 16\,GB RAM.

\subsection{nanoGPT (\texorpdfstring{$\sim$}{\textasciitilde}0.25M parameters, 4 transformer layers)}
nanoGPT is the standard workload we use for cross-project comparison. It is a
$\sim$250k-parameter GPT model with 4 transformer layers.

\paragraph{JOLT Atlas (end-to-end).}
\begin{table}[t]
  \centering
  \caption{JOLT Atlas end-to-end proving breakdown for nanoGPT.}
  \label{tab:nanogpt-jolt-atlas}
  \begin{tabular}{l r}
    \toprule
    Stage & Wall clock \\
    \midrule
    Verifying key generation & 0.246\,s \\
    Proving key generation   & 0.246\,s \\
    Proof time               & 14\,s \\
    Verify time              & 0.517\,s \\
    \bottomrule
  \end{tabular}
\end{table}

\paragraph{ezkl (same model).}
We report ezkl timings for the same model from their published benchmark.\footnote{\url{https://blog.ezkl.xyz/post/nanogpt/}}
\begin{table}[t]
  \centering
  \caption{ezkl proving breakdown for nanoGPT (reported).}
  \label{tab:nanogpt-ezkl}
  \begin{tabular}{l r}
    \toprule
    Stage & Wall clock \\
    \midrule
    Verifying key generation & 192\,s \\
    Proving key generation   & 212\,s \\
    Proof time               & 237\,s \\
    Verify time              & 0.34\,s \\
    \bottomrule
  \end{tabular}
\end{table}

\paragraph{Comparison.}
JOLT Atlas produces a proof for nanoGPT in $\sim$14\,s versus ezkl's
$\sim$237\,s proof time (not counting their $>$400\,s of key generation),
corresponding to roughly a $17\times$ speed-up on proof generation alone.

\subsection{GPT-2 (125M parameters)}
GPT-2 is a 125-million-parameter transformer model from OpenAI.

\paragraph{JOLT Atlas (end-to-end).}
\begin{table}[t]
  \centering
  \caption{JOLT Atlas end-to-end proving breakdown for GPT-2 (125M).}
  \label{tab:gpt2-jolt-atlas}
  \begin{tabular}{l r}
    \toprule
    Stage & Wall clock \\
    \midrule
    Proving/verifying key generation & 0.872\,s \\
    Witness generation              & $\sim$7.5\,s \\
    Commitment time                 & $\sim$3.5\,s \\
    Sum-check proving               & $\sim$16\,s \\
    Reduction opening proof         & $\sim$7\,s \\
    HyperKZG prove                  & $\sim$3\,s \\
    \midrule
    \textbf{End-to-end total}       & \textbf{$\sim$38\,s} \\
    \bottomrule
  \end{tabular}
\end{table}

\section{Future Work}
\label{sec:future}

Directions for future research and development include:

\begin{itemize}
    \item \textbf{Lattice-Based Polynomial Commitment Schemes:} A promising direction is replacing the existing pairing-based HyperKZG PCS with lattice-based PCS constructions to achieve post-quantum security. Several recent schemes offer compelling trade-offs:
    
    \emph{Hachi}~\cite{hachi} is a lattice-based multilinear PCS. Compared to HyperKZG, Hachi offers several advantages: (1) post-quantum security under Module-SIS assumptions, (2) faster prover time due to the absence of expensive pairing operations, and (3) $O(\sqrt{2^\ell} \cdot \lambda)$ verifier time for $\ell$-variate polynomials, approximately 12.5$\times$ faster than Greyhound~\cite{greyhound}. The main trade-off is larger proof sizes compared to pairing-based schemes.

A key consideration for lattice-based schemes is commitment size. While proof sizes are larger than pairing-based alternatives, \emph{ABBA}~\cite{abba} reduces commitment sizes through lattice-based commitment constructions built from commutators of quaternions. Smaller commitments directly benefit both proof size and verifier efficiency.
    \item \textbf{On-Chain Verification:} Deploying proofs on Ethereum is challenging due to HyperKZG's $\mathbb{G}_T$ operations. Replacing HyperKZG with a lattice-based PCS may also mitigate this limitation.
    \item \textbf{Extended ONNX Support:} Adding support for additional operators.
    \item \textbf{Adaptive teleportation for prefix--suffix efficiency.}
In the current design, TeleSparse employs a fixed global teleportation factor $\tau$ to shrink activation ranges and thereby bound lookup-table size. A natural direction for future work is to \emph{optimize} $\tau$ jointly with the prefix--suffix parameters, rather than fixing it a priori. In particular, the prefix--suffix decomposition in Eq.~\ref{eq:prefix-suffix-lookup} exposes an explicit trade-off between the number of chunks $C$, the per-chunk bit-width, and the resulting prover time and memory footprint. Since the effective activation range after teleportation directly determines the required bit-width (and hence the minimal feasible $C$), the choice of $\tau$ implicitly controls the cost profile of the lookup argument.

Future work could therefore aim to select $\tau$ so as to minimize overall proving time by balancing: (i) the streaming overhead induced by larger $C$ in Eq.~\ref{eq:prefix-suffix-lookup}, (ii) the size of the activation lookup domain after range reduction, and (iii) the induced quantization and approximation error in the model. Beyond reducing the raw range, an optimized $\tau$ may also reshape the activation input distribution (e.g., by increasing mass in saturated regions for functions such as $\tanh$), which can further improve the effectiveness of the prefix--suffix factorization by reducing the ``active'' portion of the lookup table. Developing principled methods to tune $\tau$ under explicit proof-cost and accuracy constraints remains an open and promising direction.
\end{itemize}

\vspace{1em}
The rapid development of research in both cryptography and artificial intelligence continues to provide new techniques and optimisations to integrate into Jolt Atlas. Advances in polynomial commitment schemes, folding techniques, and hardware acceleration on the cryptographic side, combined with new model architectures, quantisation methods, and inference optimisations on the AI side, ensure that verifiable machine learning will remain an active and evolving field.

\printbibliography

@string{virtual =               "Virtual Event"}

@misc{a16zbigideas,
    author = {a16z crypto},
    title = {Big Ideas in Tech 2026},
    year = {2026},
    howpublished = {\url{https://a16zcrypto.com/posts/article/big-ideas-things-excited-about-crypto-2026/}}
}

@inproceedings{ezkl,
    author = {Jason Morton and others},
    title = {EZKL: Easy Zero-Knowledge Machine Learning},
    booktitle = {GitHub Repository},
    url = {https://github.com/zkonduit/ezkl}
}

@misc{jolt,
      author = {Arasu Arun and Srinath Setty and Justin Thaler},
      title = {Jolt: {SNARKs} for Virtual Machines via Lookups},
      howpublished = {Cryptology {ePrint} Archive, Paper 2023/1217},
      year = {2023},
      url = {https://eprint.iacr.org/2023/1217}
}

@misc{joltdocs,
    author = {a16z Research},
    title = {Jolt Documentation},
    url = {https://jolt.a16zcrypto.com}
}

@misc{sumcheckisallyouneed,
      author = {Justin Thaler},
      title = {Sum-check Is All You Need: An Opinionated Survey on Fast Provers in {SNARK} Design},
      howpublished = {Cryptology {ePrint} Archive, Paper 2025/2041},
      year = {2025},
      url = {https://eprint.iacr.org/2025/2041}
}

@misc{vega,
      author = {Darya Kaviani and Srinath Setty},
      title = {Vega: Low-Latency Zero-Knowledge Proofs over Existing Credentials},
      howpublished = {Cryptology {ePrint} Archive, Paper 2025/2094},
      year = {2025},
      url = {https://eprint.iacr.org/2025/2094}
}

@misc{deepprove,
    title = {DeepProve},
    url = {https://github.com/Lagrange-Labs/deep-prove}
}

@book{thaler2022proofs,
    author = {Justin Thaler},
    title = {Proofs, Arguments, and Zero-Knowledge},
    url = {https://people.cs.georgetown.edu/jthaler/ProofsArgsAndZK}
}

@misc{telesparse,
    author = {Mohammad M. Maheri and Hamed Haddadi and Alex Davidson},
    title = {TeleSparse: Practical Privacy-Preserving Verification of Deep Neural Networks},
    year = {2025},
    howpublished = {arXiv preprint arXiv:2504.19274},
    note = {\url{https://arxiv.org/abs/2504.19274}}
}

@misc{hachi,
    author = {Ngoc Khanh Nguyen and George O'Rourke and Jiapeng Zhang},
    title = {Hachi: Efficient Lattice-Based Multilinear Polynomial Commitments over Extension Fields},
    year = {2026},
    howpublished = {Cryptology ePrint Archive, Paper 2026/156},
    note = {\url{https://eprint.iacr.org/2026/156}}
}

@misc{greyhound,
      author = {Ngoc Khanh Nguyen and Gregor Seiler},
      title = {Greyhound: Fast Polynomial Commitments from Lattices},
      howpublished = {Cryptology {ePrint} Archive, Paper 2024/1293},
      year = {2024},
      doi = {10.1007/978-3-031-68403-6_8},
      url = {https://eprint.iacr.org/2024/1293}
}

@misc{abba,
    author = {Alberto Centelles and Andrew Mendelsohn},
    title = {ABBA: Lattice-based Commitments from Commutators},
    year = {2026},
    howpublished = {Cryptology ePrint Archive, Paper 2026/148},
    note = {\url{https://eprint.iacr.org/2026/148}}
}

@misc{binius,
    author = {Benjamin E. Diamond and Jim Posen},
    title = {Succinct Arguments over Towers of Binary Fields},
    year = {2023},
    howpublished = {Cryptology ePrint Archive, Paper 2023/1784},
    note = {\url{https://eprint.iacr.org/2023/1784}}
}

@misc{bionetta,
    author = {Dmytro Zakharov and Oleksandr Kurbatov and Artem Sdobnov and Lev Soukhanov and others},
    title = {Bionetta: Efficient Client-Side Zero-Knowledge Machine Learning Proving},
    year = {2025},
    howpublished = {arXiv preprint arXiv:2510.06784}
}

@misc{groth16,
      author = {Jens Groth},
      title = {On the Size of Pairing-based Non-interactive Arguments},
      howpublished = {Cryptology {ePrint} Archive, Paper 2016/260},
      year = {2016},
      url = {https://eprint.iacr.org/2016/260}
}

@misc{ceno,
    author = {Tianyi Liu and Zhenfei Zhang and Yuncong Zhang and Wenqing Hu and Ye Zhang},
    title = {Ceno: Non-uniform, Segment and Parallel Zero-knowledge Virtual Machine},
    year = {2024},
    howpublished = {Cryptology ePrint Archive, Paper 2024/387},
    note = {\url{https://eprint.iacr.org/2024/387}}
}

@misc{hyperkzg,
    author = {Jiaxing Zhao and Srinath Setty and Weidong Cui and Greg Zaverucha},
    title = {MicroNova: Folding-based Arguments with Efficient (On-chain) Verification},
    year = {2024},
    howpublished = {Cryptology ePrint Archive, Paper 2024/2099},
    note = {\url{https://eprint.iacr.org/2024/2099}}
}

@misc{gemini,
      author = {Jonathan Bootle and Alessandro Chiesa and Yuncong Hu and Michele Orrù},
      title = {Gemini: Elastic {SNARKs} for Diverse Environments},
      howpublished = {Cryptology {ePrint} Archive, Paper 2022/420},
      year = {2022},
      url = {https://eprint.iacr.org/2022/420}
}

@article{BLUM94,
    author = {Manuel Blum and Will Evans and Peter Gemmell and Sampath Kannan and Moni Naor},
    title = {Checking the Correctness of Memories},
    journal = {Algorithmica},
    volume = {12},
    number = {2/3},
    pages = {225--244},
    year = {1994}
}

@misc{KST22,
      author = {Abhiram Kothapalli and Srinath Setty and Ioanna Tzialla},
      title = {Nova: Recursive Zero-Knowledge Arguments from Folding Schemes},
      howpublished = {Cryptology {ePrint} Archive, Paper 2021/370},
      year = {2021},
      url = {https://eprint.iacr.org/2021/370}
}

@misc{NTM25,
      author = {Vineet Nair and Justin Thaler and Michael Zhu},
      title = {Proving {CPU} Executions in Small Space},
      howpublished = {Cryptology {ePrint} Archive, Paper 2025/611},
      year = {2025},
      url = {https://eprint.iacr.org/2025/611}
}

\appendix

\end{document}